\documentclass{article}
\usepackage{spconf,amsmath,graphicx,hyperref,cite,mathtools,amsfonts,booktabs,multirow}

\newcommand{\matr}[1]{\boldsymbol{#1}}
\newcommand{\X}{\matr{X}}

\newcommand{\LA}{\matr{L}}
\newcommand{\Reg}{\matr{R}}
\newcommand{\C}{\matr{C}}

\newcommand{\vect}[1]{\boldsymbol{#1}}

\newcommand{\online}{\mathrm{on}}
\newcommand{\offline}{\mathrm{off}}

\title{Online Register for Dual-Mode Self-Supervised Speech Models: Mitigating the Lack of Future Context}
\name{Keita Goto$^{\star}$, Takashi Maekaku$^{\star}$, Jin Sakuma$^{\star}$, Jinchuan Tian$^{\dagger}$, Yusuke Shinohara$^{\star}$, Shinji Watanabe$^{\dagger}$}
\address{$^{\star}$LY Corporation, Tokyo, Japan \\
$^{\dagger}$Carnegie Mellon University, PA, USA}
\begin{document}
\ninept
\maketitle
\begin{abstract}
Dual-mode self-supervised speech models (S3Ms), which jointly pre-trained in the offline and online mode, suffer from attention mismatch in streaming scenarios due to missing future context. To address this challenge, we proposed online registers, learnable tokens appended to each chunk in online mode. These tokens act as virtual placeholders for unseen future frames, enabling the model to compensate for missing context without introducing additional latency. Furthermore, we introduce a future prediction loss that explicitly guides the registers to capture predictive cues, thereby enriching their ability to retain future information. Experiments on LibriSpeech, and out-of-domain benchmarks demonstrate that online registers consistently reduce the performance gap between offline and online modes, achieving a 3.4\% relative improvement on LibriSpeech with 160 ms chunks, especially in low-latency settings.
\end{abstract}
\section{Introduction}
\label{sec:intro}

Self-supervised speech models (S3Ms) have become an essential foundation for a wide range of speech processing systems. Typically, an S3M is pre-trained on large-scale unlabeled speech data to learn the semantic structure of speech. Various pretraining methods have been proposed, such as clustering-based targets~\cite{hubert}, random projection quantizers~\cite{best-rq,nest-rq} and differentiable vector quantization~\cite{vq-wav2vec,wav2vec}. Once pre-trained, S3Ms can be adapted to diverse downstream tasks with limited labeled data, yielding substantial accuracy improvements in automatic speech recognition (ASR)~\cite{xls-r,xeus}.

However, the representative S3Ms are still pre-trained only in offline scenarios. Unlike in the offline scenario, where the model can exploit the full utterance, the models can only access the current and past segments in online scenario. This mismatch leads to significant accuracy gaps between offline and online models.

To mitigate this issue, researchers have explored building streaming-capable S3Ms. Distillation-based methods~\cite{transducer-w2v2,streaming-w2v2,distil-w2v2} train online models to approximate the outputs of an offline model using unlabeled data. To improve robustness to chunk size, dynamic chunk training~\cite{dynamic-chunk-training} has been introduced in S3Ms~\cite{wav2vec-s}. Dual-mode ASR~\cite{dual-mode-original} unifies offline and online ASR within a single model, and UFO2~\cite{ufo2} applies this strategy to S3Ms. While these methods improve performance, a fundamental challenge remains insufficiently addressed: \textit{online models lack access to the future context that offline models can exploit}. Enlarging the chunk size or introducing look-ahead alleviates this gap, but at the cost of increased algorithmic latency.

In this work, we propose \textit{online register}, a novel approach to mitigate the lack of future context for dual-mode S3Ms. Inspired by register tokens~\cite{register-vit}, our method introduces special tokens appended to each chunk only in online mode. These tokens serve as proxies for unavailable future frames, allowing the model to exploit pseudo future context without additional latency. Moreover, by incorporating a \textit{future prediction loss} that encourages the online registers to predict future frames explicitly, the registers are guided to capture richer future information. As a result, online registers improve performance in both offline and online scenarios by effectively bridging the gap between the two modes.

\begin{figure}[t]
    \centering
    \includegraphics[width=.95\columnwidth]{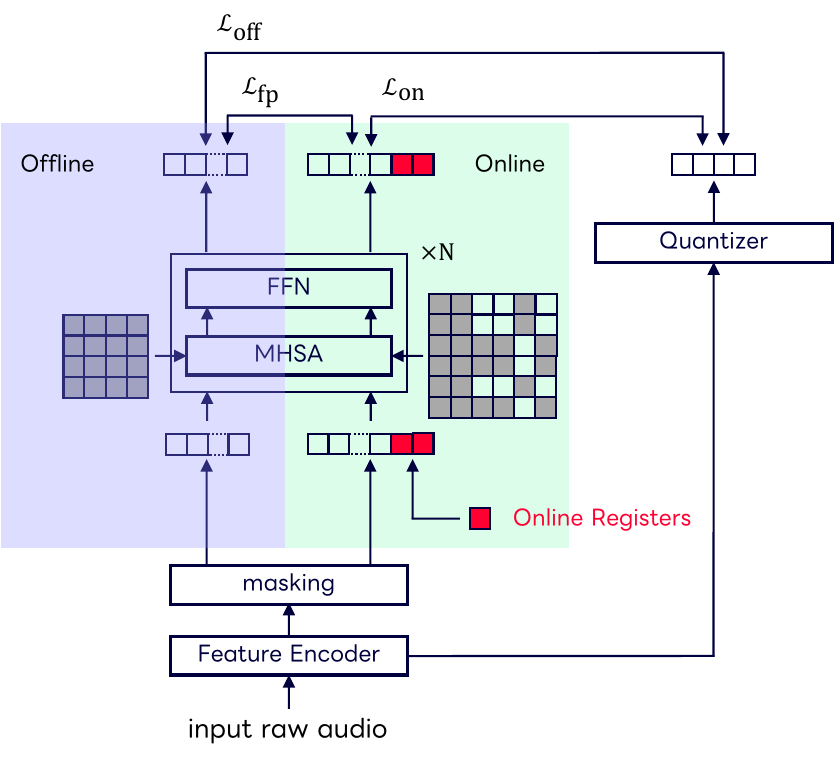}
    \vspace{-12pt}
    \caption{Overview of our proposed pre-training framework with \textit{online registers}. As an example, we illustrate the case where the feature length is 4 frames, the chunk size is 2, and the number of online registers per chunk is 1. The dual-mode Transformer encoder processes offline input with full-context attention and online input with chunk-wise attention, where online registers are appended. The model is trained to predict quantized targets for masked frames (dotted line boxes), with an additional future prediction loss encouraging the online registers to store future context.}
    \label{fig:overview}
\end{figure}

\section{Related Work}
\label{sec:related}

\noindent
\textbf{Self-supervised models}
Self-supervised speech models (S3Ms) learn useful representations from large amounts of unlabeled speech through pre-training. For example, wav2vec 2.0~\cite{wav2vec} is trained to predict quantized feature representations for masked input features, which allows the model to better capture the semantic structure of speech. Once pre-trained, S3Ms can be fine-tuned with only limited labeled data and have shown high versatility across a wide range of downstream tasks, including automatic speech recognition (ASR).

Despite these advantages, most S3Ms are pre-trained under offline full-context conditions. As a result, their performance often degrades when applied to online streaming scenarios, where only past and current speech segments are accessible. To mitigate this issue, several distillation-based approaches~\cite{transducer-w2v2,streaming-w2v2,distil-w2v2} have been proposed, training an online model to approximate the outputs of an offline model. Although these methods can effectively reduce the performance gap, they necessitate multiple training stages and thereby increase the complexity of the training pipeline.

\noindent
\textbf{Dual-mode models}
To bridge the gap between offline and online scenarios, dual-mode approaches have been proposed. Dual-mode ASR~\cite{dual-mode-original} unifies both modes within a single model by jointly training them, while UFO2~\cite{ufo2} extends this idea to self-supervised learning. In addition, dynamic chunk training~\cite{dynamic-chunk-training} is often employed in these approaches to enhance robustness against varying chunk sizes by randomly sampling chunk lengths during training.

Although these methods help narrow the performance gap, a key limitation remains: in online mode, the model cannot access future frames. Consequently, attention must be confined to the current and past chunks, whereas offline mode benefits from full-context attention. This discrepancy leads to shifted and biased attention distributions between the two modes, highlighting the need for mechanisms to compensate for missing future context without additional latency.

\noindent
\textbf{Context representations in streaming ASR}
A related line of work in streaming ASR is contextual block processing~\cite{contextual-block}, where auxiliary tokens are appended to each chunk and subsequently referenced by following chunks to transfer contextual information. Emformer~\cite{emformer} extends this idea by passing a memory vector to successive chunks, reducing the need to attend to past inputs and improving efficiency. More recently, a semi-autoregressive (SAR) streaming ASR model~\cite{semi-autoregressive} has been proposed, which incorporates previously predicted labels as additional context through a language model subnetwork. This approach improves the accuracy of streaming non-autoregressive models while maintaining low latency.

These approaches demonstrate the utility of auxiliary representations for retaining contextual information in streaming settings. However, they are not directly designed for dual-mode models, where offline and online scenarios must be handled within a single framework. In contrast, our method introduces \textit{online registers}, inspired by register tokens in vision transformers~\cite{register-vit}, to compensate for the missing future context in online mode without incurring additional latency.

\section{Proposed Method}
\label{sec:proposed}

Our method builds on the dual-mode self-supervised framework established in prior work~\cite{ufo2}, which enables a single encoder to operate under both offline and online settings. In online mode, the system sequentially produces speech representations from streaming input, whereas in offline mode, it obtains representations from the entire utterance. The novelty of our approach lies in introducing \textit{online registers}, learnable tokens that serve as virtual placeholders for missing future context in online scenarios. To further encourage these online registers to explicitly retain future information, we additionally introduce a \textit{future prediction loss}.

An overview is shown in Figure~\ref{fig:overview}. In the following subsections, we first detail the design and integration of online registers, and then describe how they are incorporated into dual-mode pre-training with the future prediction loss.

\noindent
\textbf{Online Registers.}
We introduce \textit{online registers} to mitigate the absence of future context in online scenarios. Online registers are special learnable embeddings appended to each chunk only in online mode. They can be seen as virtual placeholders for future frames. This latent workspace allows the model to accumulate predictive cues and narrow the gap between offline and online attention, simplifying the modeling of both scenarios.

Let $\X = \{\vect{x}_1, \dots, \vect{x}_T\}$ denote a sequence of $d$-dimensional vectors obtained by the convolutional feature encoder from raw speech. In a chunk-wise streaming manner, the $i$-th chunk $\C_i$ is expressed as follows:
\begin{equation}
    \label{eq:chunk}
    \C_i = \X_{(i-1)C+1:iC},
\end{equation}
where $\X_{a:b}=\{\vect{x}_a, \vect{x}_{a+1}, \dots ,\vect{x}_b\}$ and $C$ is the chunk size. For simplicity, out-of-range references $x_t$ for $t > T$ are filled with embedding vectors corresponding to padding tokens.
Optionally, look-ahead frames can be added to incorporate limited future context. With look-ahead size $L$, the $i$-th chunk has
\begin{equation}
    \label{eq:look-ahead}
    \LA_i = \X_{iC+1:iC+L}.
\end{equation}
Finally, online registers are appended to each chunk. An arbitrary number $R$ of online registers can be used, which act as learnable placeholder embeddings:
\begin{equation}
    \label{eq:register}
    \Reg_i = [\vect{r}_1, \dots, \vect{r}_R],
\end{equation}
where $\vect{r}_1, \dots, \vect{r}_R \in \mathbb{R}^d$ are learnable embedding vectors. The dimension $d$ of online registers is the same as that of the input features.

\begin{figure}[t]
    \centering
    \includegraphics[width=.7\columnwidth]{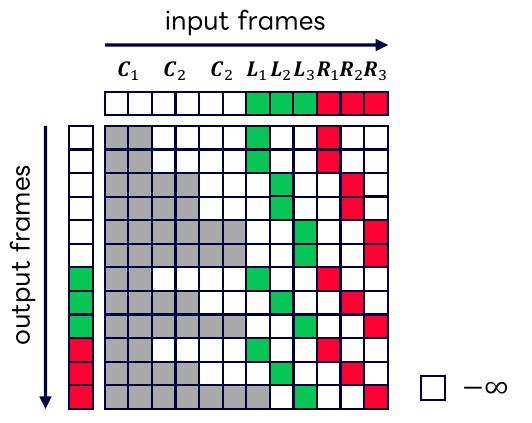}
    \vspace{-6pt}
    \caption{Attention mask design for the online mode. As an example, we illustrate the case where the feature length is 6 frames, the chunk size is 2, the look-ahead size is 1, and the number of online registers per chunk is 1. During attention computation, the model attends only to the past and current chunks, the look-ahead, and the online registers, while all other attention weights (white boxes) are filled with $-\infty$}
    \label{fig:mask}
\end{figure}

During training, all chunks, their look-ahead frames, and online registers are concatenated and fed into the encoder simultaneously:
\begin{equation}
\label{eq:online-input}
[\C_1, \C_2, \dots, \LA_1, \LA_2, \dots, \Reg_1, \Reg_2, \dots].
\end{equation}
In the Transformer encoder, an attention mask is applied to restrict the scope of attention. Figure~\ref{fig:mask} illustrates the attention mask design for the online mode. As shown in the figure, each frame is allowed to attend only to the current and past chunks, the corresponding look-ahead frames, and its associated online registers. All other attention weights are masked with $-\infty$. Unlike contextual block processing~\cite{contextual-block}, online registers are not shared across different chunks.

At inference time, the encoder instead processes input sequentially on a chunk-by-chunk basis like
\begin{equation}
[\C_1, \dots, \C_i, \LA_i, \Reg_i].
\end{equation}

\noindent
\textbf{Pre-training dual-mode S3Ms.}
We pre-train a dual-mode speech self-supervised model (S3M) with online registers. The encoder adopts a Transformer architecture and switches its attention mask depending on the mode. As shown in Figure~\ref{fig:overview}, we jointly pre-train the offline and online modes, following the wav2vec 2.0~\cite{wav2vec} framework. A feature encoder produces frame-level features from raw speech. A subset of time steps $T_\mathrm{M}$ is randomly masked, and both offline and online inputs are constructed using the same masking pattern. Online inputs are augmented with online registers at this stage as in Eq.~\ref{eq:online-input}. The dual-mode encoder computes outputs for both modes, while unmasked features are passed through a quantization module to produce target representations.

The contrastive loss is computed separately for offline and online outputs. For all masked time steps $t \in T_\mathrm{M}$, let $\vect{y}^\offline_t$ and $\vect{y}^\online_t$ denote the contextualized outputs of the dual-mode encoder in offline and online modes, respectively. Let $\vect{q}_t$ be the quantized target for step $t$, and $\matr{Q}_t$ the set consisting of $\vect{q}_t$ and $K$ distractors sampled from other masked positions of the same utterance.  
Then, the contrastive losses are defined as
\begin{equation}
\mathcal{L}_\offline = - \sum_{t \in T_\mathrm{M}} 
\log \frac{\exp(\operatorname{sim}(\vect{y}^\offline_t, \vect{q}_t)/\kappa)}
{\sum_{\vect{q}' \in \matr{Q}_t} \exp(\operatorname{sim}(\vect{y}^\offline_t, \vect{q}')/\kappa)},
\end{equation}
\begin{equation}
\mathcal{L}_\online = - \sum_{t \in T_\mathrm{M}} 
\log \frac{\exp(\operatorname{sim}(\vect{y}^\online_t, \operatorname{SG}(\vect{q}_t))/\kappa)}
{\sum_{\vect{q}' \in \matr{Q}_t} \exp(\operatorname{sim}(\vect{y}^\online_t, \operatorname{SG}(\vect{q}'))/\kappa)},
\end{equation}
where $\operatorname{sim}(\cdot,\cdot)$ denotes cosine similarity and $\kappa$ is a temperature hyperparameter.  
Following UFO2~\cite{ufo2}, we apply a stop-gradient operation $\operatorname{SG}(\cdot)$ to the targets for the online loss, so that the online representation does not affect the quantizer.  
The joint loss for the offline and online modes is
\begin{equation}
\mathcal{L}_\mathrm{dual} = \tfrac{1}{2}(\mathcal{L}_\offline + \mathcal{L}_\online) + \alpha \mathcal{L}_d,
\end{equation}
where $\mathcal{L}_d$ is the codebook diversity loss from wav2vec~2.0, and $\alpha$ is a weighting factor (set to 0.1 in our experiments).

To further encourage online registers to explicitly capture future information, we incorporate a future prediction loss $\mathcal{L}_\text{fp}$. The future prediction loss directly addresses this limitation by guiding the registers to approximate representations of future frames obtained from the offline pathway. In this way, the registers are encouraged to store richer predictive context beyond the look-ahead region, thereby narrowing the gap between online and offline modes. 

Given the output of online registers for the $i$-th chunk,
$U_i = [u^{(i)}_1, \ldots, u^{(i)}_R]$, the corresponding
future representations are obtained from the offline output $Y^\text{off}$ as
\begin{equation}
    \hat{U}_i = Y^\text{off}_{iC+L+1:iC+L+R},
\end{equation}
where $C$ is the chunk size and $L$ is the look-ahead size. Compared to Eq.~\ref{eq:chunk} and Eq.~\ref{eq:look-ahead}, note that $\hat{U}_i$ is a sequence of future representations beyond the look-ahead region.
The future prediction loss is then formulated as
\begin{equation}
    \mathcal{L}_\text{fp} = \sum_i \operatorname{MSE}(U_i, \hat{U}_i),
\end{equation}
where $\operatorname{MSE}(\cdot,\cdot)$ denotes the mean squared error.
The final training objective becomes
\begin{equation}
    \mathcal{L}_\text{dual,fp} = \mathcal{L} + \beta \mathcal{L}_\text{fp},
\end{equation}
where $\beta$ is a tunable weight. We set $\beta=1$ for our experiments.

\section{Experiments}
\label{sec:experiments}

\subsection{Experimental Settings}

\textbf{Datasets.}
We pre-trained models on LibriSpeech~\cite{librispeech} 960h without transcriptions and fine-tuned them with the official labels. \textit{dev-clean}, \textit{dev-other}, \textit{test-clean}, and \textit{test-other} were used for validation and evaluation, with checkpoints selected by validation loss on \textit{dev-other}. To assess out-of-domain performance, we additionally evaluated on the English subset of FLEURS~\cite{fleurs}.

\noindent
\textbf{Pre-training.} We pre-trained the wav2vec~2.0 BASE model, which consists of a convolutional feature encoder and 12 Transformer encoder layers. The feature encoder downsamples raw audio into frame representations with a stride of 20 ms. Following wav2vec-S~\cite{wav2vec-s}, we removed relative positional encodings and instead applied sinusoidal positional encodings. Models were initialized from the official checkpoint trained on LibriSpeech 960h.

We optimized the model and the quantizer with the Adam~\cite{adam} optimizer for 400k updates. The learning rate was warmed up for the first 32k updates to $1\times10^{-4}$ and then linearly decayed. Training was conducted with a batch size equivalent to 350 seconds of audio per GPU for about 36 hours on 16 H100 GPUs. For online mode, we adopted dynamic chunk training~\cite{dynamic-chunk-training}. For each mini-batch, the chunk size $C$ in Eq.~\ref{eq:chunk} was uniformly sampled from the range $[2, 32]$, and the look-ahead size $L$ in Eq.~\ref{eq:look-ahead} was uniformly sampled from $[0, C]$. The number of online registers $R$ in Eq.~\ref{eq:register} was fixed within the range $[1, 4]$ during pre-training. We also pre-trained a BASE model under the same settings but without online registers, serving as a baseline to fairly evaluate the effectiveness of our method. All experiments were conducted using the Fairseq framework~\cite{fairseq}.

\noindent
\textbf{Fine-tuning.} We fine-tuned the pre-trained models for ASR with the Connectionist Temporal Classification (CTC) loss~\cite{ctc}. We used LibriSpeech 960h with its transcriptions and added a randomly initialized linear projection layer on top of the Encoder to predict characters. The vocabulary consisted of 29 characters, including letters and a word boundary token.

We optimized the entire model except the feature encoder with the Adam~\cite{adam} optimizer for 320k updates. The learning rate was warmed up for the first 32k updates, held constant for the next 128k updates at $1\times10^{-4}$, and then linearly decayed. Training was conducted with a batch size equivalent to 200 seconds of audio per GPU for about 12 hours on 8 A100 GPUs. SpecAugment~\cite{specaugment} was applied to features, masking ten consecutive time steps and 64 channels with probabilities of 0.5 and 0.1, respectively. Dual-mode training and dynamic chunk training were applied with the same chunk and look-ahead ranges as in pre-training.

For LibriSpeech experiments, we decoded with the Flashlight beam search decoder~\cite{flashlight} using the officially available 4-gram language model (LM) and a beam size of 50. The LM weight and word insertion penalty were tuned with Ax~\cite{ax}, using a beam size of 50 and a fixed chunk size of 32 without look-ahead. For FLEURS, we used greedy decoding without an LM.

\subsection{Main Results}

\begin{table*}[t]
    \centering
    \caption{Word error rate (WER, \%) of baseline and proposed models on LibriSpeech \textit{test-clean}, \textit{test-other}, and FLEURS. All models are fine-tuned only on LibriSpeech 960h. All online decoding is performed with a chunk size of \textbf{160 ms} and without look-ahead. We use a single online register per chunk.}
    \label{tab:main-results}
    \begin{tabular}{@{}lllllllllll@{}}
        \toprule
        \multirow{2}{*}{Pre-train Method} & \multicolumn{2}{c}{dev-clean} & \multicolumn{2}{c}{dev-other} & \multicolumn{2}{c}{test-clean} & \multicolumn{2}{c}{test-other} & \multicolumn{2}{c}{FLEURS}\\ \cmidrule(l){2-11} 
         & Offline & Online & Offline & Online & Offline & Online & Offline & Online & Offline & Online \\ \midrule
        Dual-mode (Baseline) & 2.11 & 2.98 & 6.50 & 9.89 & 2.73 & 3.65 & 6.63 & 10.15 & 21.95 & 32.65 \\
        \quad + Online Registers & 2.08 & 2.87 & \textbf{6.41} & \textbf{9.38} & 2.70 & \textbf{3.50} & \textbf{6.52} & \textbf{9.80} & \textbf{21.72} & \textbf{32.02} \\
        \quad \quad + Future Prediction & \textbf{2.02} & \textbf{2.82} & 6.47 & 9.60 & \textbf{2.67} & 3.51 & 6.65 & 10.16 & 22.35 & 32.36 \\ \bottomrule
    \end{tabular}
    \vspace{-5mm}
\end{table*}

Table~\ref{tab:main-results} summarizes the ASR performance on LibriSpeech and FLEURS under both online and offline scenarios. We evaluate the models in a low-latency setting with a 160~ms chunk size and no look-ahead. The results confirm that online registers reduce word error rate (WER) in both modes, effectively narrowing the performance gap between online and offline modes. In contrast, the future prediction loss yields clear improvements on \textit{dev-clean}, but only marginal gains or slight degradations on other sets. We attribute this to the inherent difficulty of predicting future frames, which depends strongly on factors such as domain characteristics and chunk size. Overall, these findings highlight online registers as a robust and lightweight mechanism to compensate for missing future information without introducing additional latency.

\begin{table}[t]
    \centering
    \caption{WER (\%) of different methods on LibriSpeech \textit{test-clean} and \textit{test-other}. All online decoding is performed with a \textbf{640 ms} chunk size and without look-ahead. UFO2 does not use an external LM but instead applies attention-based rescoring.}
    \label{tab:other-methods}
    \begin{tabular}{@{}lllll@{}}
    \toprule
    \multirow{2}{*}{Method} & \multicolumn{2}{c}{test-clean} & \multicolumn{2}{c}{test-other} \\ \cmidrule(l){2-5} 
     & Offline & Online & Offline & Online \\ \midrule
    wav2vec 2.0~\cite{wav2vec} & 2.6 & - & 6.1 & - \\
    UFO2~\cite{ufo2} & 3.0 & 3.8 & 7.1 & 9.4 \\ \midrule
    Ours & 2.7 & 3.5 & 6.5 & 8.5 \\ \bottomrule
\end{tabular}
\end{table}

Table~\ref{tab:other-methods} compares our method with prior approaches. We use wav2vec 2.0~\cite{wav2vec} as an offline-only baseline and UFO2~\cite{ufo2} as a dual-mode baseline. For fairness, all systems are decoded with a 640~ms chunk size and no look-ahead. Compared to wav2vec 2.0, our model shows slightly lower performance in online mode, likely due to its dual-mode training strategy. When compared to UFO2, which is the most similar to our method, our approach achieves higher accuracy. Although decoding conditions differ across studies, our results remain competitive with those of existing methods.

\subsection{Analysis}

\noindent
\textbf{The number of register tokens.}
As shown in Table~\ref{tab:num-registers}, we analyze the relationship between the number of online registers $R$ in Eq.~\ref{eq:register} and WER. There are only marginal gains observed when increasing $R$. Notably, accuracy on the \textit{test-other} set tends to degrade as $R$ increases, with the worst result observed at $R=4$. This suggests that an excessive number of registers may cause the model to overfit to certain domains rather than generalize across diverse conditions. Considering both accuracy and computational cost, these findings indicate that using a single register is sufficient and may even be preferable in practice.

\begin{table}[t]
    \centering
    \caption{WER (\%) for different numbers of Online Registers on LibriSpeech \textit{test-clean} and \textit{test-other}. All online decoding is performed with a \textbf{160 ms} chunk size without look-ahead. All models were pre-trained without future prediction loss.}
    \label{tab:num-registers}
    \begin{tabular}{@{}lllll@{}}
        \toprule
        \multirow{2}{*}{\#Registers} & \multicolumn{2}{c}{test-clean} & \multicolumn{2}{c}{test-other} \\ \cmidrule(l){2-5} 
         & Offline & Online & Offline & Online \\ \midrule
        $R = 0$ & 2.73 & 3.65 & 6.63 & 10.15 \\
        $R = 1$ & 2.70 & 3.50 & \textbf{6.52} & \textbf{9.80} \\
        $R = 2$ & 2.73 & 3.48 & 6.70 & 10.01 \\
        $R = 3$ & \textbf{2.69} & \textbf{3.44} & 6.64 & 9.95 \\
        $R = 4$ & 2.80 & 3.56 & 6.72 & 10.19 \\ \bottomrule
    \end{tabular}
\end{table}

\noindent
\textbf{Chunk sizes.}
Table~\ref{tab:chunk_size} shows results across different chunk sizes. The chunk size $C$ determines the number of frames accessible per chunk, as defined in Eq.~\ref{eq:chunk}. Since the frame interval is 20~ms, the real-time duration of a chunk is given by $C \times 20$~ms.

We observed better performance with online registers, particularly in low-latency scenarios, such as with a 160~ms chunk size. This result suggests that online registers are especially effective when future context is severely limited. As the chunk size increased, the gap between the baseline and our method narrowed, since larger chunks implicitly provide more future context. Nevertheless, the proposed method consistently outperformed the baseline across all chunk sizes, indicating that online registers serve as a robust mechanism to mitigate the lack of future information.

Regarding the future prediction loss, its benefits are observed under limited conditions, consistent with the trends in Table~\ref{tab:main-results}. For large chunk sizes and clean datasets, modest gains are observed, likely because future prediction becomes easier in such settings. However, under more challenging conditions, performance often degrades, presumably due to the strong coupling between prediction errors and ASR errors. These observations suggest that online registers should be trained on more diverse data to generalize the benefits of future prediction more broadly.

\begin{table}[t]
    \centering
    \caption{WER (\%) for \textit{test-clean} and \textit{test-other} with different chunk sizes in online scenario. All decoding was performed without look-ahead. We use a single online register per chunk.}
    \label{tab:chunk_size}
    \begin{tabular}{@{}llll@{}}
    \toprule
    Pre-train Method & 640 ms & 320 ms & 160 ms \\ \midrule
    test-clean \\ \midrule
    Dual-mode (Baseline) & 3.21 & 3.37 & 3.65 \\
    \quad + Online Registers & 3.12 & \textbf{3.25} & \textbf{3.50} \\
    \quad\quad + Future Prediction & \textbf{3.05} & 3.27 & 3.51 \\ \midrule
    test-other \\ \midrule
    Dual-mode (Baseline) & 8.68 & 9.40 & 10.15 \\
    \quad + Online Registers & \textbf{8.48} & \textbf{9.10} & \textbf{9.80} \\
    \quad\quad + Future Prediction & 8.60 & 9.21 & 10.16 \\ \bottomrule
\end{tabular}
\end{table}

\section{Conclusion}
\label{sec:conclusion}

In this work, we introduced \textit{online registers}, a simple yet effective mechanism to compensate for the absence of future context in dual-mode self-supervised speech models. Furthermore, we proposed a \textit{future prediction loss} that explicitly encourages the registers to capture predictive cues beyond the look-ahead region. Through extensive experiments, we demonstrated that online registers improve recognition accuracy across various datasets and chunk sizes without incurring additional latency. These findings highlight the importance of lightweight auxiliary representations for bridging the offline–online gap.

\bibliographystyle{IEEEbib}
\bibliography{strings,refs}

\end{document}